# Origin and tuning of the magnetocaloric effect for the magnetic refrigerant $Mn_{1.1}Fe_{0.9}(P_{0.80}Ge_{0.20})$

Danmin Liu[1,2], Ming Yue[1] Jiuxing Zhang[1], T. M. McQueen[3], Jeffrey W. Lynn[2]*, Xiaolu Wang[1], Ying Chen[2,4], Jiying Li[2,4], R. J. Cava[3], Xubo Liu[5], Zaven Altounian[5], Q. Huang[2]

[1] Key Laboratory of Advanced Functional Materials Ministry of Education, Beijing University of Technology, 100 Pingleyuan, Chaoyang District, Beijing 100022, China
[2] NIST Center for Neutron Research, National Institute of Standards and Technology, Gaithersburg, MD 20899
[3] Department of Chemistry, Princeton University, Princeton, NJ 08544
[4] Department of Materials Science and Engineering, University of Maryland, College Park, MD 20742
[5] Center for the Physics of Materials and Department of Physics, McGill University, 3600 University Street, Montreal, Quebec, H3A 2T8, Canada

Neutron diffraction and magnetization measurements have been carried out on a series of samples of the magnetorefrigerant $Mn_{1+y}Fe_{1-y}P_{1-x}Ge_x$. The data reveal that the ferromagnetic and paramagnetic phases correspond to two very distinct crystal structures, with the magnetic entropy change as a function of magnetic field or temperature being directly controlled by the phase fraction of this first-order transition. By tuning the physical properties of this system we have achieved a magnetic entropy change (MCE) for the composition $Mn_{1.1}Fe_{0.9}P_{0.80}Ge_{0.20}$ that has a similar shape for both increasing *and* decreasing field, with the maximum MCE exceeding 74 J/kg-K—substantially higher than the previous record. The diffraction results also reveal that there is a substantial variation in the Ge content in the samples which causes a distribution of transition temperatures that reduces the MCE. It therefore should be possible to improve the MCE to exceed 100 J/kg-K under optimal conditions.

## I.    Introduction

Recently, magnetic refrigeration at ambient temperatures has attracted interest with the discovery of new materials with improved efficiencies and advantages, as potential replacements for the classical vapor compression systems in use today[1-6]. In particular, Pecharsky *et al.*[2] reported that $Gd_5(Ge_2Si_2)$ has a giant magnetocaloric effect (MCE) between 270 and 300 *K*, while Tegus *et al.*[6] found that $MnFe(P_{1-x}As_x)$ with the hexagonal $Fe_2P$-type structure has a paramagnetic-ferromagnetic phase transition that is strongly first order and exhibits a huge MCE. In addition, the Curie temperature ($T_c$) and hence optimal operating temperature of this latter material can be varied from 200 to 350 *K* by tuning the P/As ratio without losing the large MCE[6]. However, the high cost of Gd, and the toxicity of As, make it questionable whether either material will be viable commercially on a wide scale. On the other hand, recently the replacement of As by Ge or Si has been reported to still provide a very large MCE (up to 38 J/kg-K for a field change of 5 T [ref. 7-15] with Ge and up to 43 J/kg K for a field change of 3 T with Si [ref. 16]), circumventing the toxicity issue and thereby demonstrating its potential as a cost effective and environmentally friendly refrigerant[10,12-14]. For the particular optimal composition of $Mn_{1.1}Fe_{0.9}P_{0.8}Ge_{0.2}$ we report in detail here, we find that it is single phase and paramagnetic at higher temperature, single phase and ferromagnetic at lower temperature, and in between the system undergoes a strongly first-order phase transition as a function of temperature or applied magnetic field. Both phases possess the same symmetry space group ($P\bar{6}2m$) but have distinctly different structures; the *a*- and *b*-axes are ~1.3% longer while the *c*-axis is contracted by ~2.6% in the ferromagnetic (FM) phase compared to the paramagnetic (PM) phase. The large MCE of ~75 J/kg-K (see Fig. 1) on both increasing and decreasing field then originates from converting one phase to the other. The improved properties and overall advantages of this material open the possibility for its use in wide scale magnetic refrigerant applications.

## II. Experimental Procedures

The starting materials for the polycrystalline samples used in this work were submitted to ball milling, which was carried out under argon atmosphere for 1.5 hour in a high energy Pulverisette 4 mill. The milled powders were collected into a graphite mold and consolidated into a $\Phi20\times5$ mm$^3$ wafer sample at 1173 K under 30 MPa by the spark plasma sintering technique. The density of the sample was determined by the Archimedes method to be over 95% of the density of the as-cast ingot. Three 4×4×20mm bars were cut and used for the neutron diffraction experiments, but the solid bars broke into powder after measurements in which magnetic fields up to 7 T were applied together with cooling and warming between 200 *K* and 300 *K*. The results reported in this paper were obtained from the broken powder and are reproducible on cycling temperature or magnetic field.

Neutron powder diffraction (NPD) data were collected at the NIST Center for Neutron Research on the high resolution powder neutron diffractometer (BT1), with monochromatic neutrons of wavelength 1.5403 Å produced by a Cu(311) monochromator. Söller collimations before and after the monochromator and after the sample were 15′, 20′, and 7′ full-width-at-half-maximum (FWHM), respectively. Data were collected in the 2θ range of 3º to 168º with a step size of 0.05° for various temperatures from 300 K to 5 K to elucidate the magnetic and crystal structure transitions. Magnetic field measurements were carried out with a vertical field 7 T superconducting magnet. Refinements of the nuclear and magnetic structures in this system were carried out using the neutron powder diffraction data and the program GSAS[17]. The sample was found to contain about 1% MnO impurity phase that was taken into account in the refinements.

Detailed temperature and field-dependent measurements were carried out on the high-intensity BT7 and BT9 triple axis spectrometers. On each instrument a pyrolytic graphite (PG) (002) monochromator was employed to provide a neutron wavelength of 2.36 Å, and a PG filter was used to suppress higher-order wavelength contaminations. Coarse collimations of 60′, 50′, and 50′ full-width-at-half-maximum (FWHM) on BT7 and 40′, 48′, and 40′ FWHM on BT9 were employed to maximize the intensity. No analyzer was employed in these measurements.

For diffraction data obtained on BT1, a cylindrical vanadium sample holder is typically employed in order to avoid interference from diffraction peaks originating from the holder. However, we found that for temperature-dependent measurements the time to equilibrate can exceed 10-20 minutes, which can interfere with measurements of hysteresis in the present problem. We therefore undertook temperature-dependent measurements in a top-loading cryostat that avoided this problem, or used an aluminum sample holder on the BT7 and BT9 spectrometers when interference was not an issue.

Finally, we note that uncertainties where indicated in this paper are statistical in origin and represent one standard deviation.

### III. Experimental Results

Isothermal magnetization data were obtained in both magnetic field increasing and decreasing mode as shown in Fig. 1(a). The data were analyzed using the Maxwell relations method discussed in ref. [18] and [19], and the maximum entropy changes obtained are shown in Fig. 1(b). We obtained maximum magnetic entropy changes of 74 and 78 J/kg-K on increasing and decreasing field, respectively, for a field change of 5 T in the bulk sample of the $Mn_{1.1}Fe_{0.9}P_{0.8}Ge_{0.2}$. These MCE values are nearly twice the previous value for this system, and the highest MCE for any material presently reported[13,14,16].

The refined phase fraction and unit cell volume calculated from neutron diffraction refinements as a function of temperature for bulk $Mn_{1.1}Fe_{0.9}P_{0.8}Ge_{0.2}$ are shown in Fig. 2(a). The sample is in the fully paramagnetic state above ~255 K, while we find that the paramagnetic (PM) and ferromagnetic (FM) phases coexist at lower temperatures. We observed that ~80% of the sample transforms relatively quickly between 255 K and 230 K, while at 200 K the refinement shows that ~4.5% of the sample still remains in the paramagnetic phase. We show that the origin of this behavior likely is associated with variations in the Ge concentration in the sample as we discuss below. At 10 K, on the other hand, there is no detectable paramagnetic phase so that the entire sample eventually becomes fully ferromagnetic.

Figure 2(b) shows diffraction data for a temperature of 245.4 K, where 56% of the PM-phase equilibrates with 44% of the FM-phase. Only a portion of the diffraction pattern is shown for clarity. An excellent fit to the data is provided by the crystal structure shown in Fig. 3, along with a ferromagnetic structure ($P11m'$ magnetic symmetry) having Mn and Fe moments parallel in the $a$-$b$ plane. Refined ferromagnetic moments at 245 K are 2.9(1) and 0.9(1) $\mu_B$ for the Mn ($3g$) site and Fe/Mn ($3f$) site, respectively, similar to what is seen for other compounds with the $Fe_2P$-type structure where a larger moment (~3 $\mu_B$) is found at the $3g$ site and a smaller moment (<1 $\mu_B$) at the $3f$ site[10,15]. The moment direction is different from $MnFeP_{1-x}As_x$, which lies in the $a$-$c$ plane or along the $c$-axis[15], but we found that fits with a component of the moments parallel to the $c$-axis gave significantly worse agreement with our data. We remark that once the FM-phase is established, no significant changes in the crystal or magnetic structures of the FM-phase were observed on further cooling or applying a higher magnetic field. Therefore magnetic field, or temperature, has no significant effect other than to convert the system between the ferromagnetic and paramagnetic structural phases. One important finding is that there is no significant difference between the unit cell volumes of the paramagnetic and ferromagnetic phases as the phase transition proceeds.

At 295 K the entire sample is in the paramagnetic phase, so that it is single-phase and there is no magnetic order to refine. This provides the best temperature to refine the stoichiometry, and we find a Mn/Fe ratio of 1.072(6)/0.928(6) that is very close to the nominal value of Mn/Fe=1.1/0.9. We also find that the $3g$ site is completely occupied by Mn atoms, which is co-planar with P/Ge(1) atoms at the $1b$ site in the $z$=0.5 layer. The $3f$ site has ~93% Fe, with ~7% Mn distributed randomly, and the $3f$ Fe/Mn site is co-planar with the P/Ge(2) atoms at the $2c$ site in the $z$=0 layer. We also find that Ge and P are randomly mixed, although Ge atoms prefer the P/Ge(2) ($2c$) site (~27% Ge occupied) to the P/Ge(1) ($1b$) site (~5% Ge). This differs from $MnFeP_{1-x}As_x$, where Fe and Mn were found to be slightly disordered across both the $3f$ (~4% Mn) and $3g$ (~9% Fe) sites, whereas the As and P were found to have no preferential site selectivity[15]. The structural details are given in Table 1.

The first-order nature of the transition can be readily seen from magnetization data such as shown in Fig. 4(a) as a function of temperature, where the PM-phase↔FM-phase transition clearly has substantial thermal hysteresis. Since the PM-phase and FM-phase are structurally distinct, the two phases can be monitored as a function of temperature or magnetic field by neutron diffraction. For this purpose the (001) reflections of the two phases are well-resolved due to the large difference (~2.6%) in the $c$-axis, and Fig. 4(b) shows the integrated intensities for the PM-phase and FM-phase measured on cooling and warming at a speed of 15 K/hr. The peak positions are almost constant, at ~39.8° and ~41.0° in 2θ (data collected using BT9 triple axis spectrometer with a wavelength 2.36 Å) for the (001)-PM and –FM-phase, respectively, clearly indicating that this is a first order transition. We note that there was no significant difference in the hysteresis loop on repeated thermal cycling.

More importantly for the magnetic refrigerant properties, the first-order structural transition can be driven by an external magnetic field as shown in Fig. 5(a) for T = 255 K. At this temperature the diffraction data show that initially 95% of the sample is in the PM-state, which is converted into the FM-state with increasing field as indicated in Fig. 5(b) for the (001) PM-phase (nuclear only) and FM-state (nuclear + magnetic) peaks. The magnetic moments are almost independent of the field, while the unit cell volume of the FM-phase increases slightly and that of the PM-phase decreases slightly above $\mu_0 H=3$ T. The FM-phase saturates at ~5.2 T, with significant hysteresis being found on subsequently decreasing the applied field. We remark that error bars where indicated in this article are statistical in origin and represent one standard deviation.

The similar behavior of these data shows that the variation of the magnetization does in fact coincide with the FM-phase fraction. Refined phase fractions as a function of magnetic field are summarized in Fig. 6(a). The magnetic entropy change $|\Delta S_m|$ for $Mn_{1.1}Fe_{0.9}P_{0.8}Ge_{0.2}$ reported in Fig. 1 can be normalized to the magnetization, (001)-FM intensity, and fraction of FM-phase, and the corresponding values of $|\Delta S_m|$ have been inserted in Fig. 5(a), 5(b), and 6(a), respectively. The excellent agreement and linear relationship between $|\Delta S_m|$ and the FM-phase fraction (Fig. 6(b)) demonstrates that the magnetocaloric effect in the system simply mirrors the FM-phase fraction. Moreover, this investigation indicates that only ~70% of PM-phase was converted into the FM-phase in a field of 5 T. Hence an increase of $|\Delta S_m|$ up to ~100 J/kg-K could be achieved if the phase transformation goes to completion.

To further elucidate the nature of the crystal structures in both phases and how they evolve as a function of temperature and magnetic field, high-resolution neutron powder diffraction data were collected with high statistical accuracy at 245.4 K/0 T and 253.3 K/2 T. Here the PM- and FM-phases coexist, and the refinements revealed the structures for both phases in detail. Results are given in Table 1. We find that 56% of the paramagnetic phase coexists with 44% of the ferromagnetic phase at 245.4 K/0T. Interestingly, the refined average Ge occupancy for the paramagnetic portion of the sample was 0.22, almost twice the value of 0.13 for the ferromagnetic phase. We also found that at 253.3 K and a magnetic field of 2 T the portion of the sample that is in the (induced) ferromagnetic phase has a higher Ge content of 0.25 in comparison to 0.16 for the paramagnetic portion of the sample; the higher Ge content has a higher transition temperature and it is therefore easier to be converted from PM to FM by an applied magnetic field. We remark that the refinements of the two phases are relatively straightforward since the lattice parameters are quite different. We conclude that the breadth of the transition as a function of temperature or magnetic field originates primarily from the Ge inhomogeneity, which strongly suggests that it should be possible to achieve a larger MCE in a smaller applied magnetic field, if the chemical homogeneity of the sample can be improved.

## IV. Discussion

The present results directly demonstrate that the transition from the PM-phase to the FM-phase and the associated huge magnetocaloric effect is directly controlled by the first-order structural phase transition between these two phases, and, moreover, that a completed phase conversion will increase the MCE up to ~100 J/kg K in this system. The transition can be driven by temperature or applied magnetic field, and for use as a magnetic refrigerant the field-dependent properties are critical. The substitution of Ge for As removes the toxicity obstacle, which leaves the size of the required field as the primary concern. We have found that an applied field can induce substantial preferred orientation to improve the field properties, and we also expect that the magnetic properties can be optimized with further selective chemical substitutions and improved preparation techniques. Perhaps more important is the determination from the detailed crystallographic refinements that the best sample still has a substantial Ge compositional inhomogeneity, which causes a substantial spread in the transition temperature. Improvements in the compositional homogeneity should therefore permit higher MCE's to be achieved. The hysteresis associated with the transition does tend to reduce the practical MCE available for applications, but the overall improvements and prospects for further advances in performance makes this material the magnetic refrigerant of choice and should enable a wide range of commercial magnetorefrigerant applications.

In conclusion, a large magnetocaloric effect (MCE) for modest applied magnetic fields is required for a good magneto-refrigerant. The present neutron diffraction and magnetization measurements for $Mn_{1.1}Fe_{0.9}P_{0.8}Ge_{0.2}$ reveal that the ferromagnetic (FM) and paramagnetic (PM) phases correspond to two very distinct crystal structures, with the magnetic entropy change as a function of magnetic field or temperature being directly controlled by the phase fraction of this first-order transition. We find that only ~72% of the ferromagnetic phase was induced under a 5 T at the (zero-field) transition temperature, indicating that the system has a potential magnetic entropy change exceeding 100 J/kg K. Moreover, careful structural analysis reveals the presence of a substantial Ge composition inhomogeneity in the sample, suggesting that improving the chemical homogeneity is key to achieving a higher MCE.

## Acknowledgments

This work was supported by the National High Technology Research and Development Program of China (2007AA03Z458) and the Key Project of Science & Technology Innovation Engineering, Chinese Ministry of Education (705004). T. M. McQueen gratefully acknowledges support by the National Science Foundation graduate research fellowship program. The work at Princeton was supported by grant NSF-DMR-0703095. Identification of commercial equipment in the text is not intended to imply recommendation or endorsement by the National Institute of Standards and Technology.

# References

*Corresponding Author: Jeff.Lynn@nist.gov

**Figure captions**

Figure 1. (color online) (a) Isothermal magnetization curves of $Mn_{1.1}Fe_{0.9}P_{0.8}Ge_{0.2}$ on increasing and decreasing field. (b) Temperature dependence of the magnetic entropy change of the bulk $Mn_{1.1}Fe_{0.9}P_{0.8}Ge_{0.2}$ compound as a function of applied magnetic field up to 5 T.

Figure 2. (color online) (a) Phase fraction and unit cell volume as a function of temperature. About 80% of the paramagnetic phase was changed quickly to the ferromagnetic phase from 252 K to 235 K, while the remaining 20% changed slowly below 235 K. (b) Portion of the powder diffraction pattern at 245 K collected using the BT1 high resolution powder diffractometer with a wavelength 1.5403 Å. The upper panel shows a fit with only the nuclear structure model, where the difference between the observed and calculated intensities indicates the ferromagnetic peak intensities. An excellent overall fit was achieved by including a ferromagnetic model with the moments in the *a-b* plane (lower panel). The long (blue) vertical lines and the short (red) vertical lines show the structural Bragg peak positions for the ferromagnetic phase and paramagnetic phase, respectively.

Figure 3. (color online) Crystal structure with the magnetic moments (arrows) of the transition metal ions aligned ferromagnetically in the *a-b* plane.

Figure 4. (color online) (a) Temperature dependence of the magnetization in an applied field of 0.1 Tesla and rates of 15 K/hr cooling and warming. The material is not in thermodynamic equilibrium in the transition region due to hysteresis. (b) Integrated intensities of the (001) reflections for the PM- and FM- phases as a function of temperature on cooling and warming. In zero field, the magnetic moments in the FM-phase near the Curie temperature are ~3 and ~1 $\mu_B$ for the Mn and Fe/Mn sites, respectively, and increase continuously to ~3.5 and ~1.7 $\mu_B$ on cooling to 5 K.

Figure 5. (color online) (a) Field-dependent magnetization at 255 K. The magnetic entropy change $|\Delta S_m|$ (Fig. 1), normalized to the magnetization[20], is shown for comparison. (b) Field dependence of the integrated intensities of the (001) reflections for the PM-phase and FM-phase at 255 K, showing that the FM phase fraction tracks the magnetization data. For comparison, data normalized from the magnetic entropy change $|\Delta S_m|$ in Fig. 1 are also shown. The neutron data in Fig. 4b and 5b were collected using the BT7 and BT9 triple-axis spectrometers, respectively, with a wavelength 2.36 Å.

Figure 6. (a) Fraction of the ferromagnetic phase (FMP) at 255 K as the field increases. The FMP fraction increases smoothly to ~86 %. Data normalized from $|\Delta S_m|$ are shown for comparison. (b) $|\Delta S_m|$ as a function of the ferromagnetic phase fraction. The linear relationship shows that the entropy change simply tracks the FMP fraction. $|\Delta S_m|$ is projected to be ~103 J/kg-K if the transition went to completion for this sample.

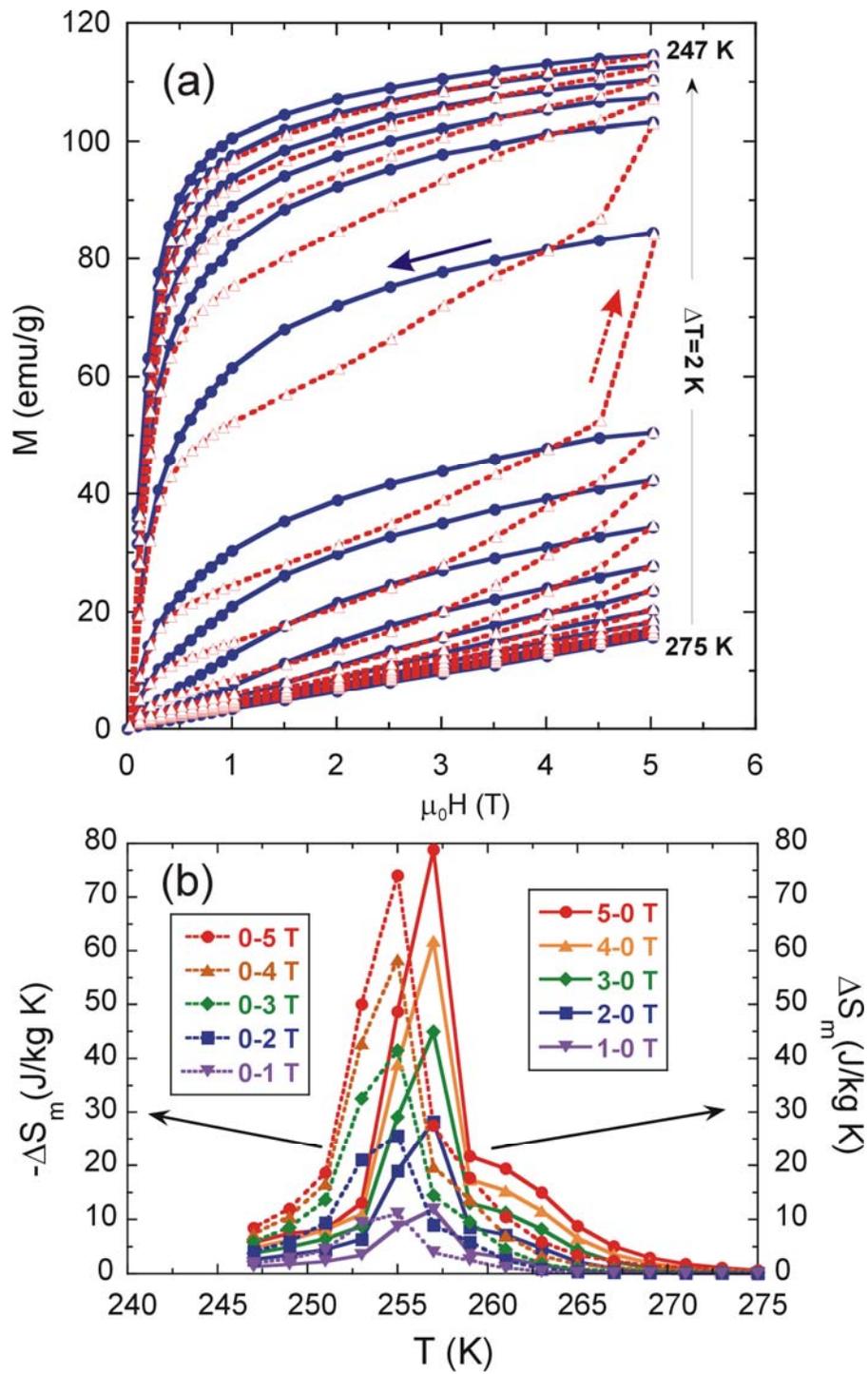

Figure 1

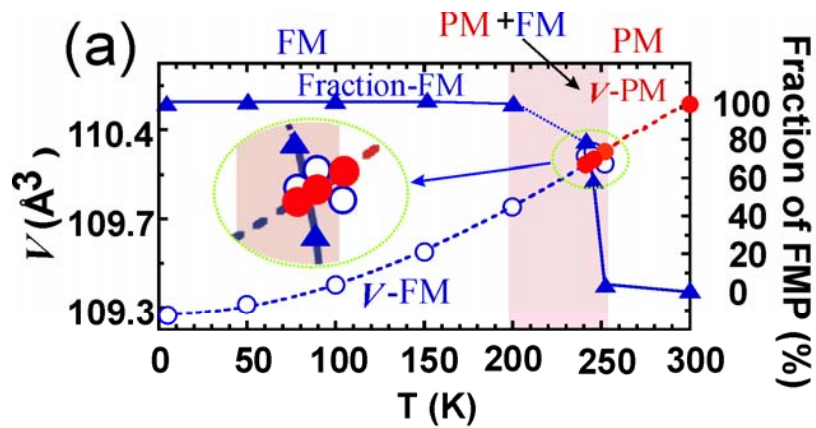

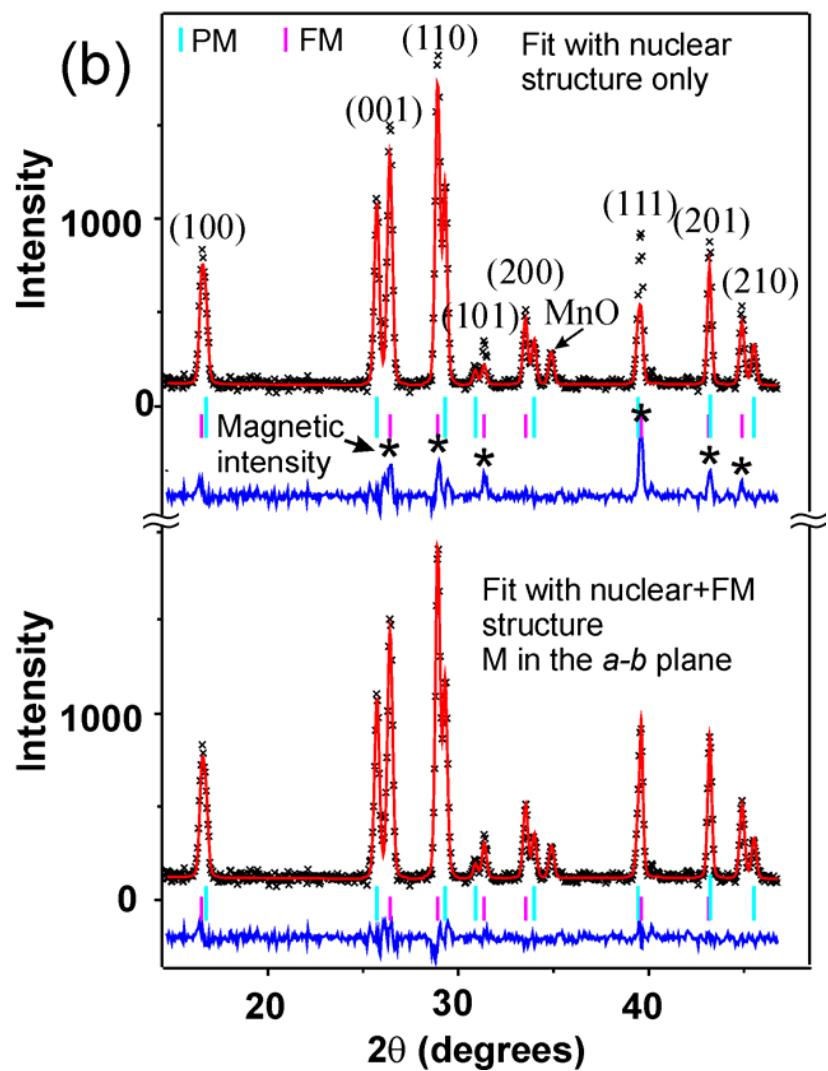

Figure 2

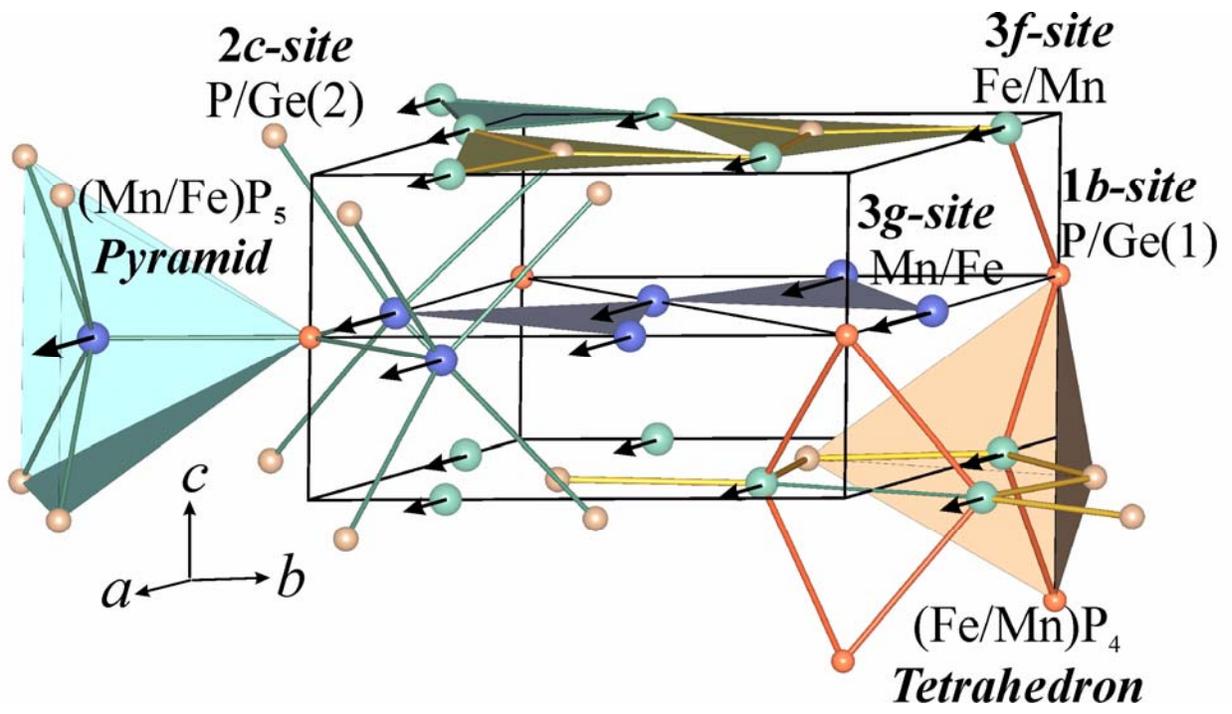

Fig. 3

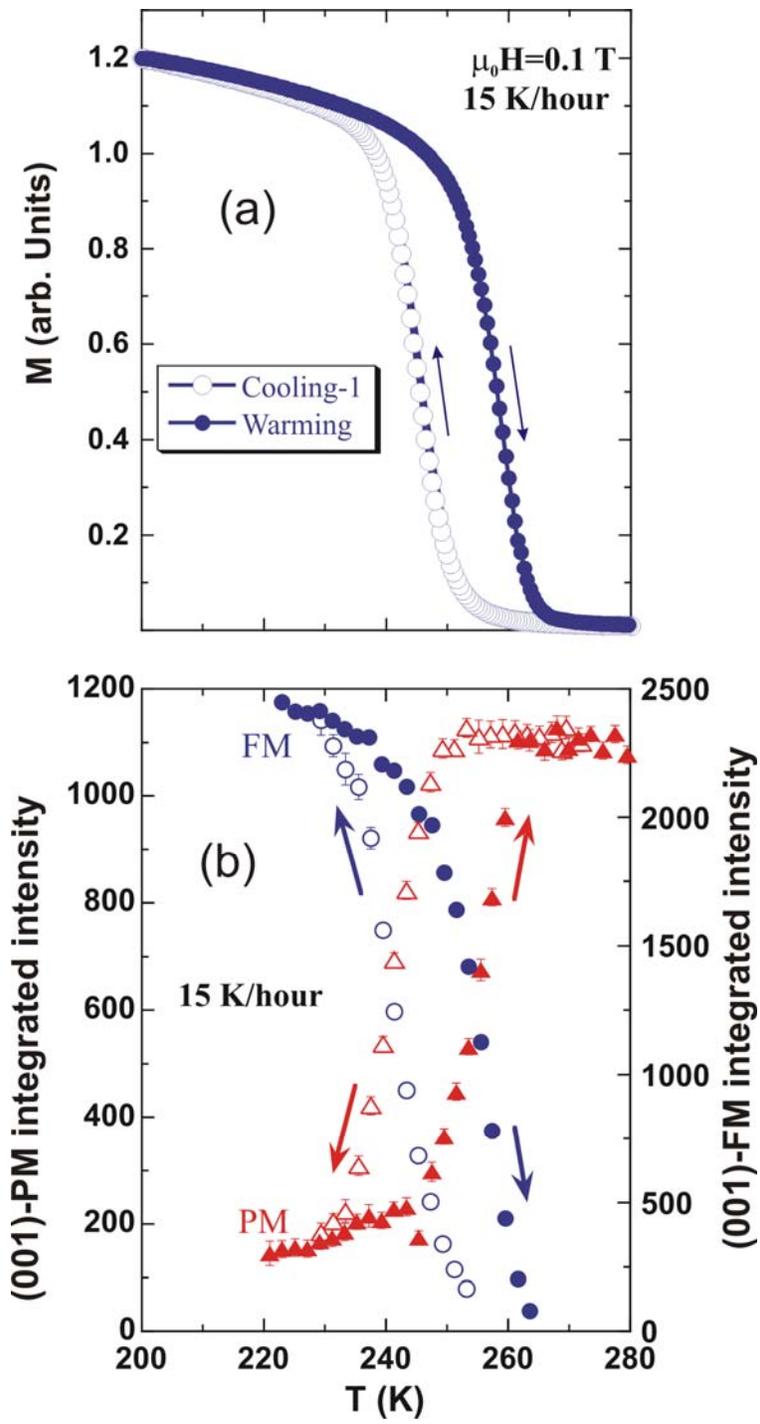

Fig. 4

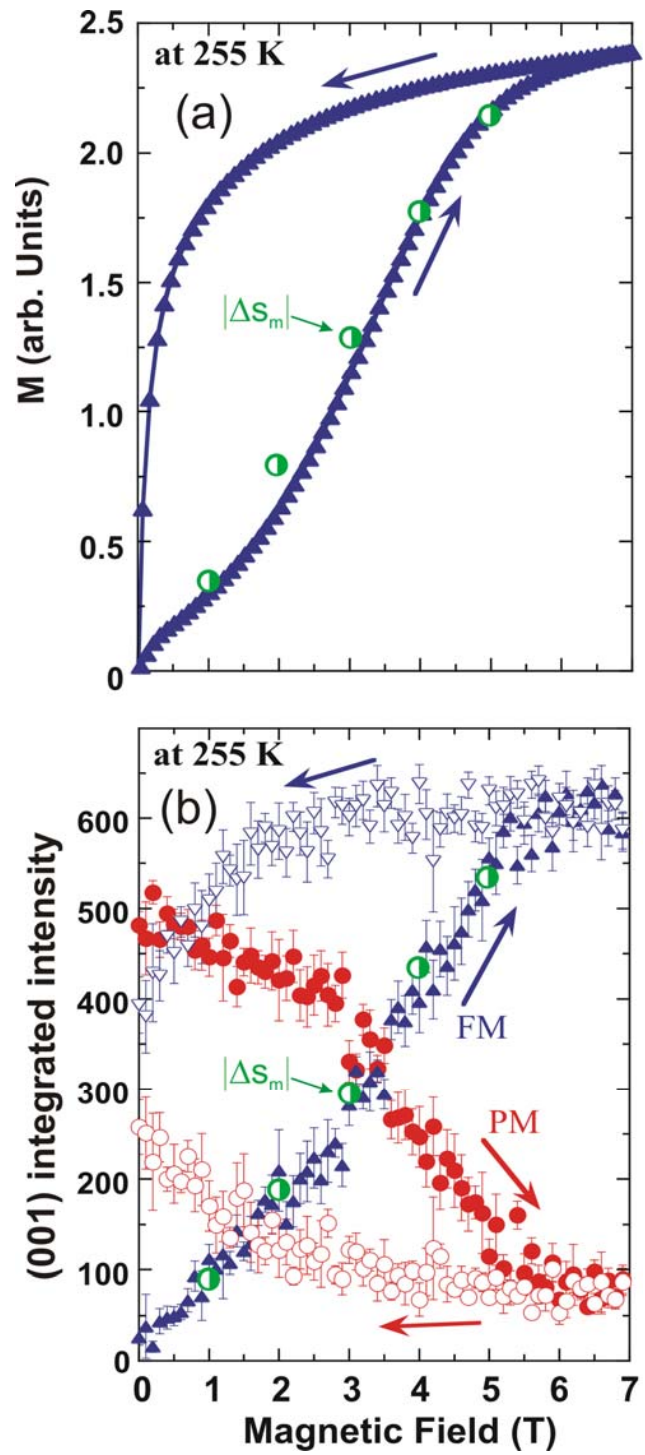

Fig. 5

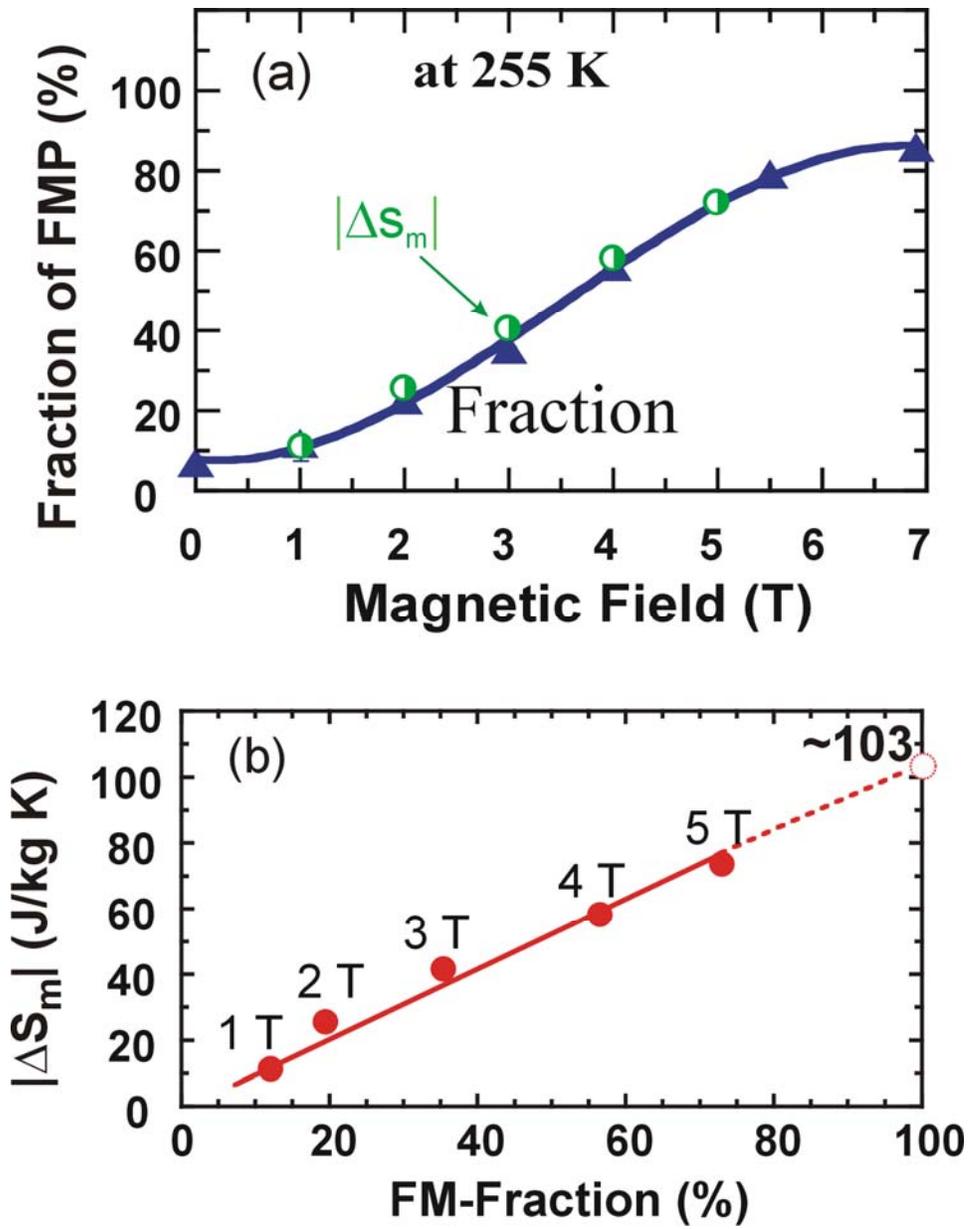

Fig. 6

Table 1. Structural parameters of $Mn_{1.1}Fe_{0.9}P_{0.8}Ge_{0.2}$ at select temperatures and magnetic fields. Space group $P\overline{6}2m$. Atomic positions: Mn: $3g(x, 0, 1/2)$; $Fe_{0.928(6)}/Mn_{0.072(6)}$: $3f(x, 0, 0)$; P/Ge(1): $1b(0, 0, 1/2)$; P/Ge(2): $2c(1/3, 2/3, 0)$. Moments for Mn and Fe were set parallel to the $a$ direction (equivalent to the $a$-$b$ plane for a powder) in the refinements.

| Atom | Parameters | 295 K/0 T | 245.4 K/0 T | | 200 K/0 T | 253.3 K/2 T | |
|---|---|---|---|---|---|---|---|
| | | PM-phase | PM-phase | FM-phase | FM-phase | PM-phase | FM-phase |
| Refined Fraction | | 100% | 56.0(1)% | 44.0(1)% | 95.6(3)% | 66.7(1)% | 33.1(1)% |
| Refined $n$(P)/$n$(Ge) | | 0.80/0.20 | 0.78/0.22 | 0.87/0.13 | 0.84/0.16 | 0.84/0.16 | 0.75/0.25 |
| | $a$ (Å) | 6.06137(7) | 6.0705(1) | 6.1515(1) | 6.1605(1) | 6.0698(1) | 6.1496(2) |
| | $c$ (Å) | 3.46023(5) | 3.4490(1) | 3.3592(1) | 3.33822(9) | 3.4522(1) | 3.3637(1) |
| | $V$ (Å$^3$) | 110.098(3) | 110.070(3) | 110.084(3) | 109.718(3) | 110.149(3) | 110.164(5) |
| Mn/Fe | $x$ | 0.5916(3) | 0.5929(5) | 0.5974(8) | 0.5949(1) | 0.600(1) | 0.600(1) |
| | $B$(Å$^2$) | 0.77(2) | 0.79(2) | 0.79(2) | 0.56(2) | 1.20(3) | 1.20(3) |
| | $n$(Mn)/$n$(Fe) | 0.996/0.004(2) | 0.990/0.010(2) | 0.984/0.016(3) | 0.989/0.011(4) | 0.987/0.013(3) | 0.979/0.021(5) |
| | $M$ (μ$_B$) | | | 2.9(1) | 2.9(1) | | 3.4(1) |
| Fe/Mn | $x$ | 0.2527(1) | 0.2534(2) | 0.2546(3) | 0.2558(2) | 0.2546(5) | 0.2546(5) |
| | $B$(Å$^2$) | 0.77(2) | 0.79(4) | 0.79(4) | 0.56(2) | 1.20(3) | 1.20(3) |
| | $n$(Fe)/$n$(Mn) | 0.930/0.070(2) | 0.912/0.088(3) | 0.907/0.093(3) | 0.916/0.084(4) | 0.921/0.079(3) | 0.955/0.045(7) |
| | $M$ (μ$_B$) | | | 0.9(1) | 1.4(1) | | 0.9(1) |
| P/Ge(1) | $B$(Å$^2$) | 0.55(4) | 0.70(4) | 0.70(4) | 0.57(4) | 0.90(6) | 0.90(6) |
| | $n$(P)/$n$(Ge) | 0.92/0.08(1) | 0.90/0.10(2) | 0.96/0.04(2) | 0.95/0.05(1) | 0.99/0.01(3) | 0.84/0.16(4) |
| P/Ge(2) | $B$(Å$^2$) | 0.55(4) | 0.70(4) | 0.70(4) | 0.57(4) | 0.90(6) | 0.9(6) |
| | $n$(P)/$n$(Ge) | 0.74/0.26(1) | 0.72/0.28(2) | 0.83/0.17(2) | 0.79/0.21(1) | 0.76/0.24(2) | 0.71/0.29(3) |
| | $Rp$ (%) | 5.25 | 3.19 | | 7.98 | 5.45 | |
| | $wRp$ (%) | 6.65 | 5.14 | | 10.01 | 6.96 | |
| | $\chi^2$ | 1.276 | 1.972 | | 1.498 | 1.886 | |